# Spatiotemporal Nonlinear Pulse Dynamics in Multimode Silicon Nitride Waveguides

A?ka M?ula I?kandar M?da, and U?ur Te?in*

*Electrical and Electronics Engineering, Koç University, Istanbul, Türkiye*
*utegin@ku.edu.tr*

**Abstract:** We present an open-source multimode nonlinear Schrödinger equation-based simulation to investigate spatiotemporal nonlinear pulse propagation in thin-film silicon nitride (SiN) waveguides. Using this framework, we analyze femtosecond pulse evolution under diverse excitation conditions in a 6-µm-wide SiN waveguide supporting six TE modes. Our results reveal that mode selection and power distribution critically govern nonlinear coupling, soliton fission, and dispersive wave generation, leading to broadband spectra exceeding 3 µm. Our findings reveal that input mode engineering is a powerful strategy for tailoring ultrafast nonlinear dynamics in integrated photonic platforms, with applications in supercontinuum generation, frequency combs, and programmable nonlinear optics.

## 1. Introduction

Silicon nitride (SiN) has emerged as one of the most versatile material platforms for nonlinear integrated optics [1]. Its relatively large Kerr nonlinearity, wide transparency window, and compatibility with CMOS fabrication have enabled various demonstrations, from frequency comb generation to broadband supercontinuum sources [2–6]. Compared to silicon, SiN exhibits lower two-photon absorption in the near-infrared and a smaller refractive index contrast with silica cladding, mitigating modal walk-off and facilitating phase matching across multiple spatial modes [1,7]. These advantages make SiN particularly attractive for extending nonlinear optics beyond the single-mode regime into the complex landscape of multimode spatiotemporal nonlinear dynamics.

Spatiotemporal nonlinear optics introduces an additional spatial degree of freedom to explore complex propagation dynamics [8]. In multimode fibers, this has led to the observation of remarkable phenomena, including spatiotemporal instability [9,10], beam self-cleaning [11,12], multimode solitons [13], dispersive waves [14], high-power supercontinuum generation [15–17], and spatiotemporal mode-locking [18–20]. These effects arise from intermodal interactions and nonlinear coupling mechanisms that enrich propagation dynamics far beyond what is achievable in single-mode systems. Recent advances have further demonstrated that spatiotemporal interactions can be controlled—and even exploited for computing tasks—by combining linear and nonlinear propagation with machine learning techniques [21–24]. Such results highlight that spatiotemporal nonlinear optics is both a fertile ground for fundamental science and a promising pathway for novel applications.

Inspired by these discoveries in fibers, recent research has focused on integrated multimode SiN waveguides, where strong light confinement and engineered dispersion create new opportunities. On-chip demonstrations include mode-selective four-wave mixing [25], intermodal parametric processes for wideband generation [26], intermodal dispersive wave emission [27], and octave-spanning supercontinuum generation [28]. These pioneering works confirm that multimode SiN waveguides exhibit a similarly rich variety of spatiotemporal dynamics as fibers, offering unprecedented control through waveguide design and mode-selective excitation [29]. However, systematic studies on how excitation conditions and mode superpositions govern nonlinear pulse propagation in SiN remain limited.

In this work, we address this gap by developing an open-source simulation framework based on the multimode nonlinear Schrödinger equation (MM-NLSE) to investigate spatiotemporal nonlinear dynamics in SiN waveguides. Focusing on an 800 nm thick, 6 µm wide waveguide supporting multiple TE0X modes up to 2.2 µm, we simulate the propagation of a 250 fs Gaussian pulse with 25 kW peak power under a range of excitation conditions. By varying the power distribution among modes, from fundamental-only excitation to combinations including higher-order modes, we examine how symmetry, mode order, and energy allocation influence nonlinear processes such as self-phase modulation, intermodal coupling, soliton fission, and dispersive wave emission. Our results reveal distinct signatures of multimode nonlinear propagation and establish a foundation for engineering broadband spectra and spatiotemporal control in integrated photonic platforms.

## 2. Methods

Spatiotemporal nonlinear propagation in multimode waveguides has been studied extensively with fiber optics in recent decades, and numerical tools for studying the underlying complex dynamics to generate nonlinear dynamics have matured significantly [8,30]. Here we utilize the same numerical approach and develop an open source library for integrated multimode waveguides.

Our numerical tool is based on the multimode nonlinear Schrödinger equation (MM-NLSE). MM-NLSE relies on decomposing the input field into the waveguide modes and calculating the propagation of light in each mode with coupled nonlinear Schrödinger equations. For each mode $A_p$ propagation is governed by

$$\frac{\partial A_p(z,t)}{\partial z} = i(\beta_0^{(p)} - \beta_0)A_p(z,t) - (\beta_1^{(p)} - \beta_1)\frac{\partial A_p(z,t)}{\partial t} + i\sum_{n\geq 2}\frac{\beta_n^{(p)}}{n!}\left(i\frac{\partial}{\partial t}\right)^n A_p(z,t)$$
$$+ i\frac{n_2\omega_0}{c}\sum_{l,m,n}Q_{plmn}(\omega_0)A_l(z,t)A_m(z,t)A_n^*(z,t)$$

where $Q_{plmn}$ is nonlinear coupling coefficient, $n_2$ is the nonlinear Kerr coefficient and $\omega_0$ is the central frequency and $\beta_n^{(p)}$ is Taylor expansion coefficient of propagation constant for corresponding mode. The nonlinear coupling tensor ($Q_{plmn}$) which dictates the intermodal coupling coefficients between the modes is calculated using the spatial overlap integral

$$Q_{plmn} = \frac{\iint F_p(x,y)F_l(x,y)F_m(x,y)F_n(x,y)dxdy}{[\iint|F_p|^2 dxdy \iint|F_l|^2 dxdy \iint|F_m|^2 dxdy \iint|F_n|^2 dxdy]^{1/2}}$$

where $F_p(x,y)$ is the transverse field of modes. For each calculation step, nonlinear propagation is calculated with state-of-the-art 4th order Runge-Kutta in the interaction picture algorithm to maintain highest numerical accuracy [31].

The waveguide considered in our study has 800 nm thickness and 6000 nm width. Such a waveguide supports at least 6 modes up to 2.2 μm wavelength when the refractive index distribution is analyzed by a quasi-vectorial mode solver [32]. The calculated modes and the their respective dispersion curves are presented in Fig. 1. Here, we would like to highlight that the fundamental (TE$_{00}$) mode has two zero dispersion wavelength (ZDW) at 1006 nm and 1696 nm. However, at 1550 nm the second order dispersion coefficient ($\beta_2^{(p)}$) is anomalous across all modes.

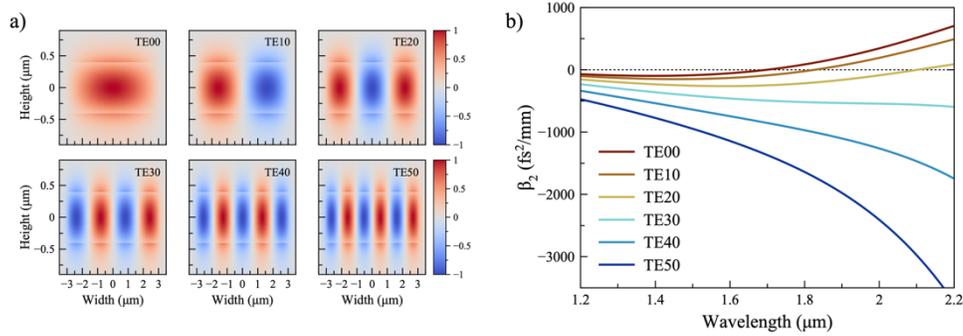

Fig. 1. Mode function and dispersion curves of the silicon nitride waveguide. (a) Transverse Electric (TE) Mode functions, calculated at 1550 nm. (b) Dispersion curves between 1.2 $\mu m$ and 2.2 $\mu m$ wavelength range.

After defining the modes of waveguide, nonlinear mode coupling coefficients are calculated. The effective area was found to be approximately 3.36 μm$^2$ for the fundamental transverse mode TE$_{00}$. To visualize the intermodal coupling dynamics, the nonlinear coupling tensor ($Q_{plmn}$) is flattened and normalized as shown in the Fig. 2. Our results indicate that the intermodal mixing of four waves between modes TE$_{00}$ and TE$_{20}$, and between modes TE$_{10}$ and TE$_{50}$ are the largest. In this waveguide odd and even TE$_{x0}$ modes tend to mix between odd and even modes, respectively.

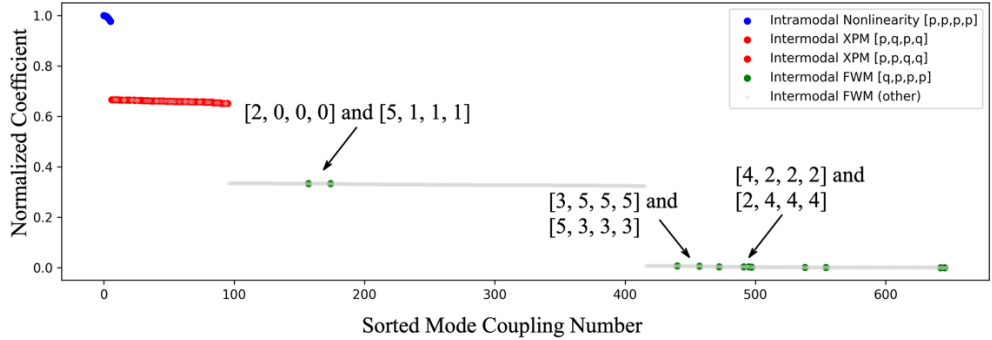

Fig. 2. Flattened intermodal coupling coefficient tensor.

In our simulations, time grid of $2^{16}$ across 50 ps time window is utilized to accommodate possible modal walk-off between the modes. We implement a super Gaussian boundary to prevent pulse from going across time window. Raman effect is not considered, since silicon nitride is known to exhibit low Raman fraction has no major contribution to supercontinuum formation in silicon nitride waveguides [3,33].

## 3. Results and Discussion

We select the input pulse with a Gaussian temporal shape, 250 fs duration and 25 kW peak power at 1550 nm central wavelength. The length of SiN waveguide is defined as 3 mm to provide sufficient length for spatiotemporal nonlinear mixing between the waveguide modes. To understand the spatiotemporal nonlinear dynamics of the femtosecond pump pulse in the multimode thin-film Si$_3$N$_4$ waveguide, we perform a systematic investigation under various excitation conditions.

## 3.1. Single mode excitation

When the pump pulse is launched into the fundamental transverse mode (TE$_{00}$) of the waveguide, the nonlinear dynamics are constrained by the symmetry of the excitation as shown in Fig. 3. We observe that strong self-phase modulation rapidly broadens the TE$_{00}$ spectrum, and after 2.5 mm of propagation a broadband supercontinuum spanning from 1 μm to 3.3 μm emerges. During the propagation, power remains confined within the even-order modes, with sequential transfer from TE$_{00}$ to TE$_{20}$, and subsequently to TE$_{40}$. As dictated by the nonlinear coupling dynamics, coupling from fundamental transverse mode to odd-order modes does not take place.

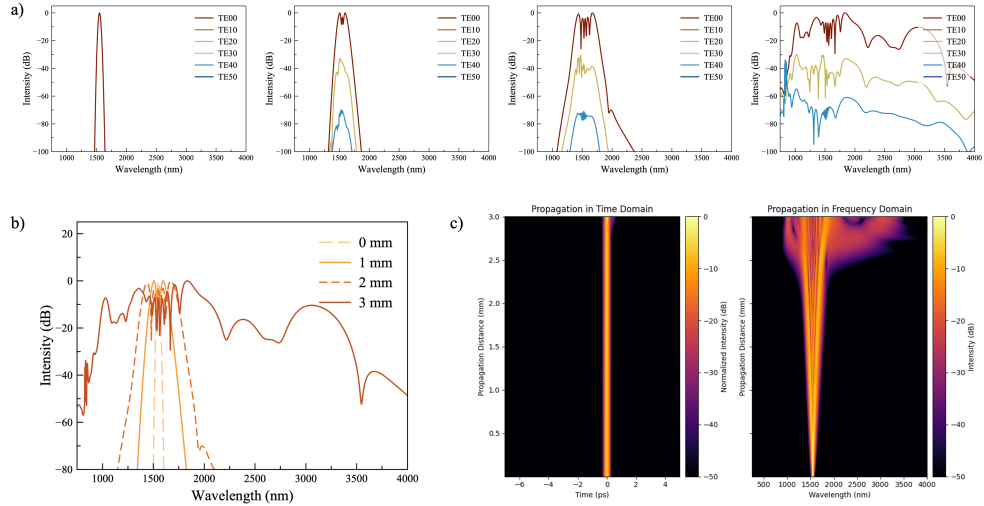

Fig. 3. Overview of multimode pulse evolution in time and frequency domains under TE$_{00}$ excitation. (a) Spectral evolution of each TE mode. From left to right: 0 mm, 1 mm, 2 mm, 3 mm propagation. (b) Temporal evolution of pulse for each TE mode. From left to right: 0 mm, 1 mm, 2 mm, 3 mm propagation. (c) Total output spectra at different propagation lengths. (d) Time- and frequency-domain evolution during propagation.

## 3.2. Mode-pair excitations

We next examine mode-pair excitation scenarios by equally dividing the 25 kW peak power between two guided modes. First, we consider simultaneous excitation of the TE$_{00}$ and TE$_{20}$ modes. This configuration leads to richer nonlinear dynamics and more intricate spectral broadening compared to the single-mode case. As shown in Fig. 4, multimode interactions remain confined to even-order modes. Owing to its shorter nonlinear length, TE 20 experiences stronger spectral broadening than TE$_{00}$. After 2 mm of propagation, a supercontinuum extending up to 4 μm emerges, albeit with reduced flatness relative to the single-mode scenario.

Next, when the input power is equally distributed between TE$_{00}$ and TE$_{10}$, the excitation symmetry is broken. This triggers nonlinear coupling across both even- and odd-order modes, enabling all six guided modes to participate in the evolution. The resulting spectrum is broader and noticeably flatter than in previous cases (Fig. 5). Notably, TE$_{10}$ enhances long-wavelength broadening, pushing the spectrum beyond 3 μm.

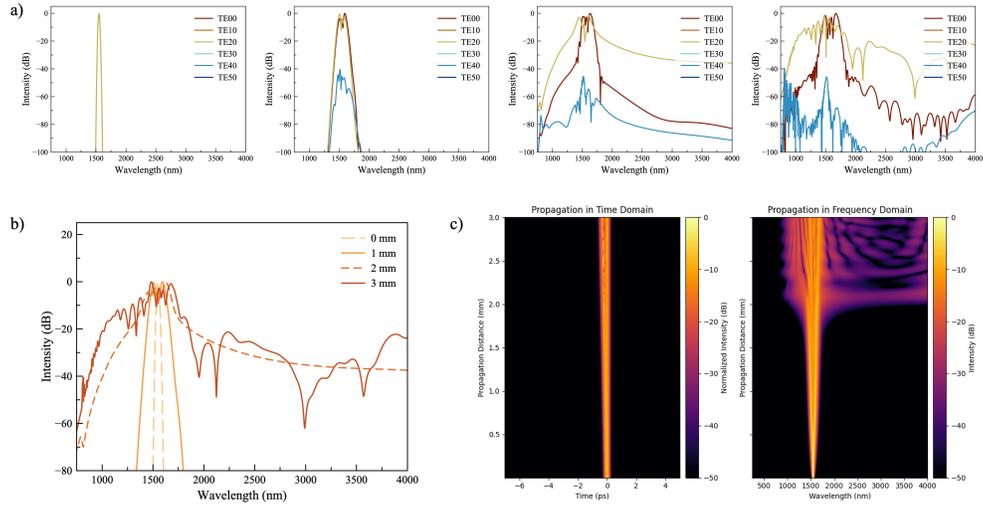

Fig. 4. Overview of multimode pulse evolution in time and frequency domains under $TE_{00}$ and $TE_{20}$ excitation. (a) Spectral evolution of each TE mode. From left to right: 0 mm, 1 mm, 2 mm, 3 mm propagation. (b) Temporal evolution of pulse for each TE mode. From left to right: 0 mm, 1 mm, 2 mm, 3 mm propagation. (c) Total output spectra at different propagation lengths. (d) Time- and frequency-domain evolution during propagation.

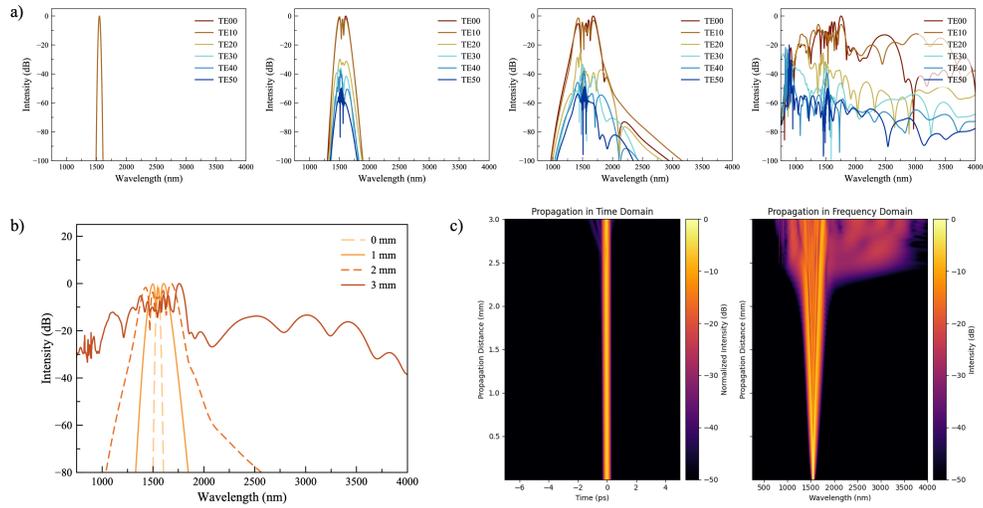

Fig. 5. Overview of multimode pulse evolution in time and frequency domains under $TE_{00}$ and $TE_{10}$ excitation. (a) Spectral evolution of each TE mode. From left to right: 0 mm, 1 mm, 2 mm, 3 mm propagation. (b) Temporal evolution of pulse for each TE mode. From left to right: 0 mm, 1 mm, 2 mm, 3 mm propagation. (c) Total output spectra at different propagation lengths. (d) Time- and frequency-domain evolution during propagation.

Finally, we investigate an extreme case involving simultaneous excitation of $TE_{00}$ and $TE_{40}$. Due to the large modal index contrast, this pair exhibits distinct dynamics (Fig. 6). Strong intermodal walk-off causes temporal separation of the pulses associated with $TE_{00}$ and $TE_{40}$. Within the $TE_{40}$ branch, soliton fission occurs after approximately 1.5 mm of propagation, generating a pronounced dispersive wave near 800 nm. These observations highlight the interplay between nonlinear interactions and intermodal group-velocity mismatch.

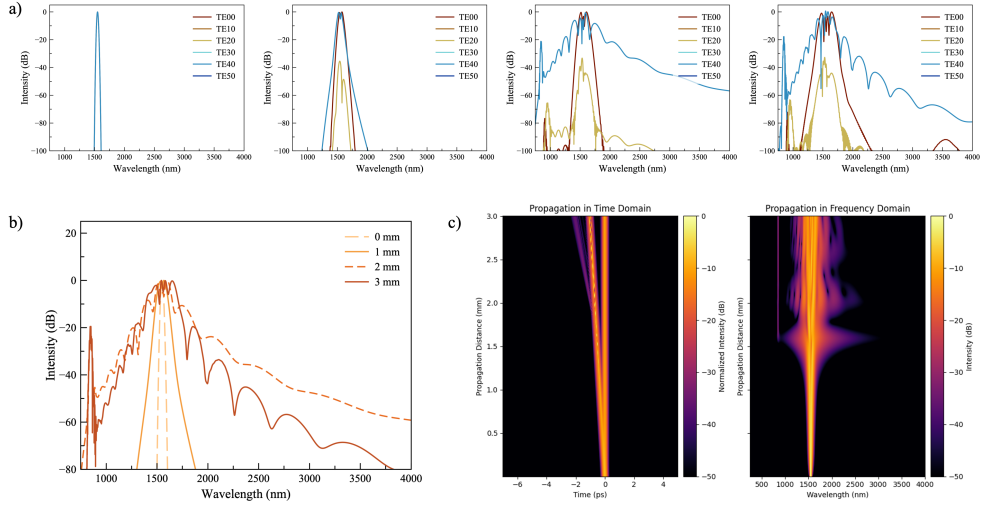

Fig. 6. Overview of multimode pulse evolution in time and frequency domains under $TE_{00}$ and $TE_{40}$ excitation. (a) Spectral evolution of each TE mode. From left to right: 0 mm, 1 mm, 2 mm, 3 mm propagation. (b) Temporal evolution of pulse for each TE mode. From left to right: 0 mm, 1 mm, 2 mm, 3 mm propagation. (c) Total output spectra at different propagation lengths. (d) Time- and frequency-domain evolution during propagation.

### 3.3. Distributed excitations

We then investigate distributed excitation scenarios. When the input power is evenly distributed across all six guided modes, each mode carries approximately 4.17 kW, substantially reducing nonlinear strength. Under these conditions, temporal walk-off dominates the propagation, leading to near-complete separation of the modal pulses. Consequently, the output spectrum remains relatively narrow, exhibiting only mild broadening compared to previous cases. This result emphasizes that distributing power too evenly suppresses nonlinear interactions and limits continuum formation.

A markedly different behavior is observed when most of the input power (70%) is concentrated in the highest-order mode, $TE_{50}$, with progressively smaller fractions in lower-order modes (20% to $TE_{40}$, 10% to $TE_{30}$, 5% to $TE_{20}$, 2.5% to $TE_{10}$, 2.5% to $TE_{00}$). The high peak power in $TE_{50}$ drives very early soliton fission, occurring after only 0.8 mm of propagation. This rapid breakup produces a highly structured output spectrum dominated by short-wavelength components. A strong dispersive wave emerges near 900 nm from $TE_{50}$, while $TE_{40}$ contributes a secondary peak around 800 nm. These features highlight the sensitivity of multimode supercontinuum generation to the initial power distribution, demonstrating how strategic allocation can drastically alter the resulting spectrum.

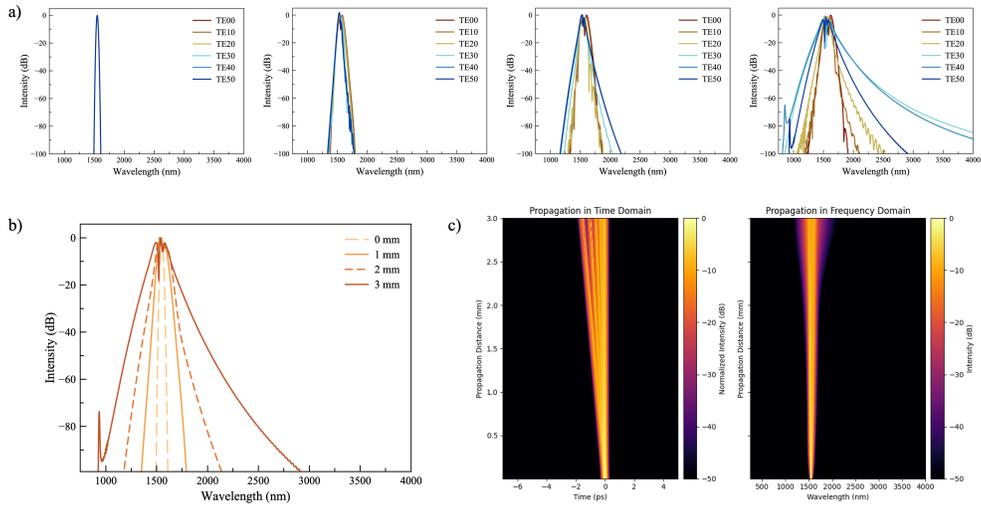

Fig. 7. Overview of multimode pulse evolution in time and frequency domains when pulse energy is equally distributed between the modes. (a) Spectral evolution of each TE mode. From left to right: 0 mm, 1 mm, 2 mm, 3 mm propagation. (b) Temporal evolution of pulse for each TE mode. From left to right: 0 mm, 1 mm, 2 mm, 3 mm propagation. (c) Total output spectra at different propagation lengths. (d) Time- and frequency-domain evolution during propagation.

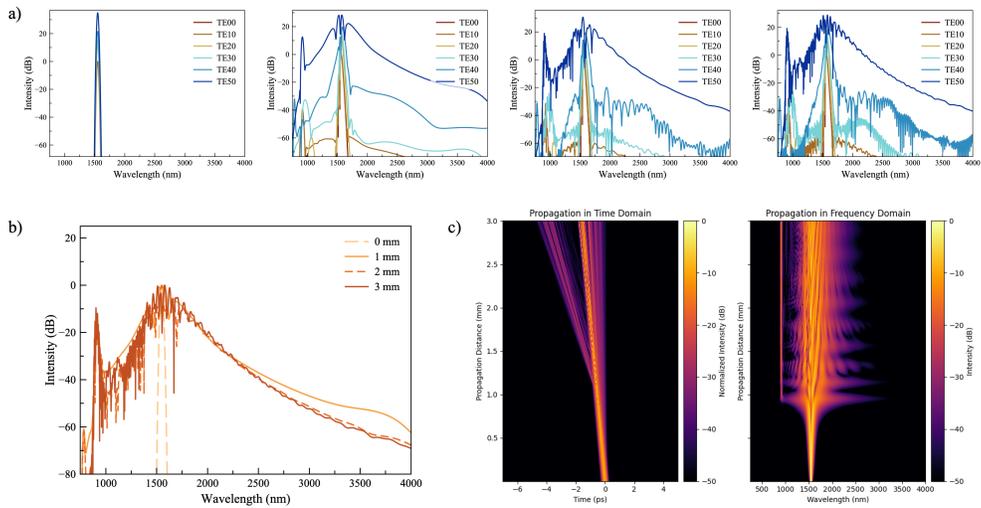

Fig. 8. Overview of multimode pulse evolution in time and frequency domains when high order mode excitation is preferred. (a) Spectral evolution of each TE mode. From left to right: 0 mm, 1 mm, 2 mm, 3 mm propagation. (b) Temporal evolution of pulse for each TE mode. From left to right: 0 mm, 1 mm, 2 mm, 3 mm propagation. (c) Total output spectra at different propagation lengths. (d) Time- and frequency-domain evolution during propagation.

Our results underscore the pivotal role of excitation symmetry and energy distribution in governing nonlinear dynamics in multimode silicon nitride waveguides. When excitation is limited to even modes, power exchange remains confined within that symmetry class, restricting four-wave mixing pathways and limiting modal diversity. Introducing odd modes breaks this constraint, unlocking cross-parity interactions and enabling a richer coupling landscape and propagation. This expanded interaction space supports more uniform power exchange, resulting in broader and flatter spectral profiles.

Energy placement across modes also proves critical. Concentrating power in higher-order modes accelerates soliton fission and drives dispersive wave emission at shorter wavelengths, key mechanisms for generating broadband supercontinua. In contrast, distributing power too evenly across all modes diminishes nonlinear efficiency. The resulting dominance of temporal walk-off suppresses spectral broadening, which, while limiting bandwidth, may be advantageous in applications requiring spectral stability.

## 4. Conclusion

We present an open-source multimode nonlinear Schrödinger equation-based simulation to study spatiotemporal nonlinear pulse propagation in silicon nitride (SiN) and investigate spatiotemporal nonlinear dynamics across a broad range of excitation conditions. Implementing Kerr nonlinearity with the Finite Difference Time Domain (FDTD) method is challenging [34]. In FDTD method-based calculations, solving the nonlinearity with perturbative and iterative methods using the explicit method would result in instability in the simulation. Sophisticated solutions such as Alternating Direction Implicit (ADI) FDTD or Hybrid Implicit-Explicit scheme can mitigate the instability up to certain intensity levels [35,36]. However, extreme nonlinear events such as supercontinuum generation are still challenging to study with the FDTD method. Our simulation offers a solution to study nonlinear dynamics in integrated waveguides.

By systematically varying the excited modes and their power distributions, we identified how symmetry and modal overlap shape pulse evolution through mechanisms such as self-phase modulation, intermodal energy transfer, soliton fission, and dispersive wave generation. Our analysis highlights the central role of parity in determining modal interactions. Excitation restricted to even modes confines nonlinear coupling within that branch, whereas introducing odd modes breaks these constraints and enables broader modal participation. Furthermore, power distribution across modes strongly influences the dynamics: concentrating energy in higher-order modes accelerates soliton fission. It enhances short-wavelength dispersive wave emission, while uniform power sharing amplifies temporal walk-off, suppressing nonlinear efficiency and limiting continuum formation.

These insights establish input mode engineering as a powerful strategy for tailoring ultrafast nonlinear processes in integrated photonic platforms. By carefully selecting mode combinations and power allocation, one can realize desired spatiotemporal behaviors, from broad and flat supercontinuum to structured spectra with targeted dispersive features. This capability offers new opportunities for designing compact, chip-scale light sources with controllable spectral characteristics, with applications spanning supercontinuum generation, mid-infrared frequency combs, programmable nonlinear optics, and neuromorphic photonic computing.

**Funding.** This work is supported by Optica Foundation Challenge Prize.

**Disclosure.** The authors declare no conflicts of interest.

**Data availability.** The results presented in this paper are obtained via our simulation tool which is available in Github Page (https://github.com/utegin-lpt/Integrated-MMNLSE).